\documentclass[aps,reprint,superscriptaddress,onecolumn,notitlepage,longbibliography]{revtex4-1}
\pdfoutput=1

\usepackage{graphicx}
\usepackage{dcolumn}
\usepackage{bm}
\usepackage{bbm}
\usepackage{color}
\usepackage{url}
\usepackage{multirow} 
\usepackage[normalem]{ulem}
\usepackage{xcolor}
\usepackage{makecell}
\usepackage{amsmath}
\usepackage[colorlinks=true, urlcolor=blue, citecolor=blue, anchorcolor=red]{hyperref}
\usepackage{algorithm}
\usepackage{setspace}
\usepackage[noend]{algpseudocode}
\usepackage[section]{placeins} 

\newcommand{\be}{\begin{equation}}
\newcommand{\ee}{\end{equation}}

\newcommand{\sgn}{{\rm sgn}}

\newcommand{\btheta}{{\mbox{\boldmath $\theta$}}}

\newcommand{\bphi}{{\mbox{\boldmath $\phi$}}}

\newcommand{\tr}{\mathrm{tr}}

\begin{document}

\title{Adversarial quantum circuit learning for pure state approximation}

\author{Marcello Benedetti}
\email{m.benedetti@cs.ucl.ac.uk}
\affiliation{Department of Computer Science, University College London, WC1E 6BT London, United Kingdom}
\affiliation{Cambridge Quantum Computing Limited, CB2 1UB Cambridge, United Kingdom}

\author{Edward Grant}
\affiliation{Department of Computer Science, University College London, WC1E 6BT London, United Kingdom}
\affiliation{Rahko Limited, NN14 1UQ Northamptonshire, United Kingdom}

\author{Leonard Wossnig}
\affiliation{Department of Computer Science, University College London, WC1E 6BT London, United Kingdom}
\affiliation{Rahko Limited, NN14 1UQ Northamptonshire, United Kingdom}

\author{Simone Severini}
\affiliation{Department of Computer Science, University College London, WC1E 6BT London, United Kingdom}

\date{April 15, 2019}

\begin{abstract}
Adversarial learning is one of the most successful approaches to modelling high-dimensional probability distributions from data. The quantum computing community has recently begun to generalize this idea and to look for potential applications. In this work, we derive an adversarial algorithm for the problem of approximating an unknown quantum pure state. Although this could be done on universal quantum computers, the adversarial formulation enables us to execute the algorithm on near-term quantum computers. Two parametrized circuits are optimized in tandem: One tries to approximate the target state, the other tries to distinguish between target and approximated state. Supported by numerical simulations, we show that resilient backpropagation algorithms perform remarkably well in optimizing the two circuits. We use the bipartite entanglement entropy to design an efficient heuristic for the stopping criterion. Our approach may find application in quantum state tomography.
\end{abstract}

\maketitle

\section{Introduction}

In February $1988$ Richard Feynman wrote on his blackboard: `What I cannot create, I do not understand'~\cite{feynman1988}. Since then this powerful dictum has been reused and reinterpreted in the context of many fields throughout science. In the context of machine learning, it is often used to describe generative models, algorithms that can generate realistic synthetic examples of their environment and therefore are likely to `understand' such an environment. 

Generative models are algorithms trained to approximate the joint probability distribution of a set of variables, given a dataset of observations. Conceptually, the quantum generalization is straightforward; Quantum generative models are algorithms trained to approximate the wave function of a set of qubits, given a dataset of quantum states. This process of approximately reconstructing a quantum state is already known to physicists under the name of quantum state tomography. Indeed, there already exist proposals of generative models for tomography such as the quantum principal component analysis~\cite{lloyd2014quantum} and the quantum Boltzmann machine~\cite{amin2018quantum,kieferova2017tomography}. Other machine learning approaches for tomography have been formulated using the different framework of probably approximately correct learning~\cite{aaronson2007learnability,rocchetto2017experimental}. Hence, machine learning could provide a new set of tools to physicists. Going the other way, quantum mechanics could provide a new set of tools to machine learning practitioners for tackling classical tasks. As an example, Born machines~\cite{cheng2018information,benedetti2018generative} use the probabilistic interpretation of the quantum wave function to reproduce the statistics observed in classical data. Identifying classical datasets that can be modelled better via quantum correlations is an interesting open question in itself~\cite{perdomo2018opportunities}.

One of the most successful approaches to generative models is that of adversarial algorithms in which a discriminator is trained to distinguish between real and generated samples, and a generator is trained to confuse the discriminator~\cite{goodfellow2014generative}. The intuition is that if a generator is able to confuse a perfect discriminator, then it means it can generate realistic synthetic examples. Recently, researchers have begun to generalize this idea to the quantum computing paradigm~\cite{dallaire2018quantum,lloyd2018quantum} where the discriminator is trained to distinguish between two sources of quantum states. The discrimination of quantum states is so important that it was among the first problems ever considered in the field of quantum information theory~\cite{helstrom1969quantum}. The novelty of adversarial algorithms is in using the discriminator's performance to provide a learning signal for the generator. 

But how do generative models stand with respect to state-of-the-art algorithms already in use on quantum hardware? The work on the variational quantum eigensolver~\cite{peruzzo2014variational} shows that parametrized quantum circuits can be used to extract properties of quantum systems, e.g., the electronic energy of molecules. Similarly, the work on quantum approximate optimization~\cite{farhi2014quantum} shows that parametrized quantum circuits can be used to obtain good approximate solutions to hard combinatorial problems, e.g., the max-cut. All these problems consist of finding the ground state of a well-defined, task-specific, Hamiltonian. However, in generative models the problem is somehow inverted. We ask the question: What is the Hamiltonian that could have generated the statistics observed in the dataset? Although some work has been done in this direction~\cite{verdon2017quantum,benedetti2018generative}, much effort is required to scale these models to a relevant size. Moreover, it would be preferable for models to make no unnecessary assumption about the data. These are the aspects where we expect adversarial quantum circuit learning to stand out.

Notably, adversarial quantum circuits do not perform quantum state tomography in the strict sense, since the entries of the target density matrix are never read out explicitly. Instead, they perform an \textit{implicit} state tomography by learning the parameters of the generator circuit, i.e., an implicit description of the resulting state. This approach does hence not suffer from the exponential cost incurred by the long sequence of adaptive measurements required in standard state tomography. This is because, as we will see, only one qubit needs to be measured in order to train and adapt the circuit. The subtlety here is that an exponential cost could occur through a non-converging training process. However, we did not observe this in practice. Our results also allow for a range of potential applications, which we detail below.

As a first example of interest to physicists, one can use the approach to find a Tensor Network representation of a complex target state. In this scenario, the structure of the generator circuit is set up as a Tensor Network and the method learns its parameters. The only assumption here is that the target state can be loaded to the quantum computer via a physical interface with the external world. As a second example of interest to computer scientists, one can use the approach to `compile' a known sequence of gates to a different or simpler sequence. In this scenario, the target is the state generated by the known sequence of gates, and the generator is the `compiled' circuit. This could have concrete applications such as the translation of circuits from superconducting to ion trap gate sets.

In this manuscript, we start from information theoretic arguments and derive an adversarial algorithm that learns to generate approximations to a target pure quantum state. We parametrize generator and discriminator circuits similarly to other variational approaches, and analyze their performance with numerical simulations. Our approach is designed to make use of near-term quantum hardware to its fullest extent, including for the estimation of the gradients necessary to learn the circuits. Optimization is performed using an adaptive gradient descent method known as resilient backpropagation~(Rprop)~\cite{riedmiller1993direct}, which performs well when the error surface is characterized by large plateaus with small gradient, and only requires that the sign of the gradient can be ascertained. We provide a heuristic method to assess the learning, which can in turn be used to design a stopping criterion. Although our simulations are carried out in the context of noisy intermediate-scale quantum computers~(NISQ)~\cite{preskill2018quantum}, we discuss long-term realizations of the adversarial algorithm on universal quantum computers.

\section{Method}\label{s:method}

Consider the problem of generating a pure state $\rho_g$ close to an unknown pure target state $\rho_t$, where closeness is measured with respect to some distance metric to be chosen. Hereby we use subscripts $g$ and $t$ to label `generated' and `target' states, respectively. The unknown target state is provided a finite number of times by a channel. If we were able to learn the state preparation procedure, then we could generate as many `copies' as we want and use these in a subsequent application. We now describe a game between two players whose outcome is an approximate state preparation for the target state.

Borrowing language from the literature of adversarial machine learning, the two players are called the generator and the discriminator. The task of the generator is to prepare a quantum state and fool the other player into thinking that it is the true target state. Thus, the generator is a unitary transformation $G$ applied to some known initial state, say $|0\rangle$, so that ${ \rho_g = G |0\rangle \langle 0|G^\dag }$. We will discuss the generator's strategy later. 

The discriminator has the task of distinguishing between the target state and the generated state. It is presented with the mixture ${\rho_{mix} =  P(t) \rho_t + P(g) \rho_g}$, where $P(t)$ and $P(g)$ are prior probabilities summing to one. Note that in practice the discriminator sees one input at a time rather than the mixture of density matrices, but we can treat the uncertainty in the input state using this picture. The discriminator performs a positive operator-valued measurement~(POVM) $\{E_b\}$ on the input, so that $\sum_b E_b = I$. According to Born's rule, measurement outcome $b$ is observed with probability $P(b)=\tr[E_b \rho_{mix}]$. The outcome is then fed to a decision rule, a function that estimates which of the two states was provided in input.

A straightforward application of Bayes' theorem suggests that the decision rule should select the label for which the posterior probability is maximal, i.e., $\arg\max_{x \in \{g, t\}} P(x|b)$. This rule is called the Bayes' decision function and is optimal in the sense that, given an optimal POVM, any other decision function has a larger probability of error~\cite{fuchs1996distinguishability}. Recalling that $\max_{x \in \{g, t\}} P(x|b)$ is the probability of the correct decision using the Bayes decision function, we can formulate the probability of error as
\be
\begin{split}
P_{err}(\{E_b\}) &= \sum_{b} P(b) (1 - \max_{x \in \{g, t\}} P(x|b))\\
&=\sum_{b} P(b) \min_{x \in \{g, t\}} P(x|b) \\
&=\sum_{b} \min_{x \in \{g, t\}} P(x|b) P(b) \\
&=\sum_{b} \min_{x \in \{g, t\}} P(b|x) P(x) \\
&=\sum_{b} \min_{x \in \{g, t\}} \tr[E_b \rho_x] P(x) .
\end{split}
\label{e:bayes}
\ee
We observe that the choice of POVM plays a key role here; the discriminator should consider finding the best possible one. Therefore, we can write the objective function for the discriminator in variational form as
\be
P_{err}^* = \min_{\{E_b\}} P_{err}(\{E_b\}) ,
\label{e:optimal_d}
\ee
where the minimization is over all possible POVM elements, and the number of POVM elements is unconstrained.

It was Helstrom who carefully designed a POVM achieving the smallest probability of error when a single sample of $\rho_{mix}$ is provided~\cite{helstrom1969quantum}. He showed that the optimal discriminator comprises two elements, $E_0$ and $E_1$, which are diagonal in a basis that diagonalizes ${\Gamma = P(t) \rho_t - P(g) \rho_g}$. When the outcome $0$ is observed, the state is labeled as `target', when the outcome $1$ is observed the state is labeled as `generated'. This is the discriminator's optimal strategy as it minimizes the probability of error in Eq.~\ref{e:optimal_d}. Unfortunately, designing such a measurement would require knowledge of the target state beforehand, contradicting the purpose of the game at hand. Yet we now know that the optimal POVM comprises only two elements. Using this information, and plugging Eq.~\eqref{e:bayes} in Eq.~\eqref{e:optimal_d}, we obtain~\cite{fuchs1996distinguishability}
\be
\begin{split}
P_{err}^* &= \min_{\{E_0, E_1\}} \Big ( P(1|t) P(t ) + P(0|g) P(g) \Big ) \\
&= \min_{\{E_0, E_1\}} \Big ( \tr [ E_1 \rho_t ] P(t)  +  \tr [ E_0 \rho_g ] P(g) \Big ) \\
&= \hspace{.71em} \min_{E_0} \hspace{.71em} \Big ( \tr [ (I - E_0) \rho_t ] P(t)  +  \tr [ E_0 \rho_g ] P(g) \Big )\\
&= \hspace{.71em} \min_{E_0} \hspace{.71em} \Big ( - \tr [ E_0 \rho_t ] P(t)  +  \tr [ E_0 \rho_g ] P(g) \Big ) + P(t),
\end{split}
\label{e:prob_err}
\ee
where we used $E_1 = I - E_0$ from the definition of POVM. We now return to the generator and outline its strategy. Assuming the discriminator be optimal, the generator achieves success by maximizing the probability of error $P_{err}^*$ with respect to the generated state $\rho_g$. The result is a zero-sum game similar to that of generative adversarial networks~\cite{goodfellow2014generative} and described by
\be
\begin{split}
&\max_{\rho_g} \min_{E_0} \Big ( - \tr [ E_0 \rho_t ] P(t)  +  \tr [ E_0 \rho_g ] P(g) \Big ) \\
=& \min_{\rho_g} \max_{E_0} \Big ( \tr [ E_0 \rho_t ] P(t) - \tr [ E_0 \rho_g ] P(g) \Big ) ,
\end{split}
\label{e:minimax}
\ee
where we dropped the constant terms. Now suppose that the game is carried out in turns. On the one side, the discriminator is after an unknown Helstrom measurement which changes over time as the generator plays. On the other side, the generator tries to imitate an unknown target state exploiting the signal provided by the discriminator. 

Note that when $P(t) = P(g) = \frac{1}{2}$, the probability of error in Eq.~\eqref{e:optimal_d} is related to the trace distance between quantum states~\cite{nielsen2011quantum}
\be
\begin{split}
D(\rho_t, \rho_g) &= \frac{1}{2} \| \rho_t - \rho_g \| \\
&= \max_{\{E_b\}} \frac{1}{2} \sum_b \big | \tr [E_b (\rho_t - \rho_g) ]  \big | .
\end{split}
\label{e:tracedist}
\ee
This is clearer from the variational definition in the second line. Hence, by playing the minimax game above with equal prior probabilities, we are implicitly minimizing the trace distance between target and generated state. We will use the trace distance to analyze the learning progress in our simulations. In practice though, one does not have access to the optimal POVM in Eq.~\eqref{e:tracedist}, because that would require, once again, the Helstrom measurement. We discuss this ideal scenario in Section~\ref{s:error_corrected} where we require the availability of a universal quantum computer. We shall now consider the case of implementation in NISQ computers where, due to the infeasibility of computing Eq.~\eqref{e:tracedist}, we need to design a heuristic for the stopping criterion.

Finally, we note that this game, based on the Bayesian probability of error, assumes the availability of one copy of $\rho_{mix}$ at each turn. A more general minimax game could be designed based on the quantum Chernoff bound assuming the availability of multiple copies at each turn~\cite{audenaert2007discriminating,fuchs1996distinguishability}.

\subsection{Near-term implementation on NISQ computers}

We now discuss how the game could be played in practice using noisy quantum computers and no error correction. First, we assume the ability to efficiently provide the unknown target state as an input. In realistic scenarios, the target state would come from an external channel and would be loaded in the quantum computer's register with no significant overhead. For example, the source may be the output of another quantum computer, while the channel may be a quantum internet.

Second, the generator's unitary transformation shall be implemented by a parametrized quantum circuit applied to a known initial state. Note that target and generated states have the same number of qubits and they are never input together, but rather as a mixture with probabilities $P(t)$ and $P(g)$, respectively, i.e., randomly selected with a certain prior probability. Hence they can be prepared in the same quantum register. 

Third, resorting to Neumark's dilation theorem~\cite{neumark1940spectral}, the discriminator's POVM shall be realized as a unitary transformation followed by a projective measurement on an extended system. Such extended system consists of the quantum register shared by the target and generated states, plus an ancilla register initialized to a known state. Notice that the number of basis states for the ancillary system needs to match the number of POVM elements. Because here we specifically require two POVM elements, the ancillary system consists of just one ancilla qubit. The unitary transformation on this extended system is also implemented by a parametrized quantum circuit. The measurement is described by projectors on the state space of the ancilla and the two possible outcomes, 0 and 1, are respectively associated with labels `target' and `generated'. 

Depending on the characteristics of the circuits, such as type of gates, depth, and connectivity, we will be able to explore regions of the Hilbert space with the generator, and explore regions of the cone of positive operators with the discriminator.

As a concrete example, assume that the unknown $n$-qubit target state $\rho_t = |\psi_t\rangle\langle \psi_t|$ is prepared in the main register $\mathcal{M}$. We construct a generator circuit ${G = G_L \cdots G_1}$ where each gate is either fixed, e.g. a CNOT, or parametrized. Parametrized gates are often of the form ${G_l(\theta_l)=\exp(-i \theta_l H_l/2)}$ where $\theta_l$ is a real valued parameters and ${H_l \in \{X, Y, Z, I\}^{\otimes n}}$ is a tensor product of $n$ Pauli matrices. The generator acts on the initial state ${|0\rangle^{\otimes n}}$ and prepares ${\rho_g = G|0\rangle\langle 0|G^\dag}$ in the main register $\mathcal{M}$. We then similarly construct a discriminator circuit ${D = D_K \cdots D_1}$ acting non-trivially on both main register $\mathcal{M}$ and ancilla qubit $\mathcal{A}$. Each gates is either fixed or parametrized as ${D_k(\phi_k)=\exp(-i \phi_k H_k / 2)}$, where $\phi_k$ is real valued and $H_k$ is a tensor product of $n+1$ Pauli matrices. We measure the ancilla qubit using projectors ${ E_b = I^{\otimes n} \otimes |b\rangle\langle b| }$ with $b \in \{0, 1\}$. Collecting parameters for generator and discriminator into vectors $\btheta$ and $\bphi$, respectively, the minimax game in Eq.~\eqref{e:minimax} can be written as ${ \min_{\btheta} \max_{\bphi} V(\btheta, \bphi) }$ with value function
\be
\begin{split}
V(\btheta, \bphi) =& \tr \Big [ E_0  D \Big ( |\psi_t\rangle\langle \psi_t| \otimes |0\rangle\langle 0| \Big ) D^\dag \Big ] P(t) - \tr \Big [ E_0  D \Big ( G|0\rangle\langle 0|G^\dag \otimes |0\rangle\langle 0| \Big ) D^\dag \Big ] P(g).
\end{split}
\label{e:value_function}
\ee
Each player optimizes the value function in turn. This optimization can in principle be done via different approaches (e.g., gradient-free, first-, second-order methods, etc.) depending on the computational resources available. Here we discuss a simple method of alternated optimization by gradient descent/ascent starting from randomly initialized parameters $\btheta^{(0)}$ and $\bphi^{(0)}$. That is, we perform iterations of the form ${\btheta^{(t+1)} = \arg\min_{\btheta} V(\btheta, \bphi^{(t)})}$ and ${\bphi^{(t+1)} = \arg\max_{\bphi} V(\btheta^{(t+1)}, \bphi)}$. 

To start with, we need to compute the gradient of the value function with respect to the parameters. The favorable properties of the tensor products of Pauli matrices appearing in our gate definitions allow computation of the analytical gradient using the method proposed in Ref.~\cite{mitarai2018quantum}. For the generator, the partial derivatives read  
\be
\begin{split}
\frac{\partial V}{ \partial \theta_l} = -\frac{P(g)}{2} \Big \{ &  \tr \Big [ E_0 D \Big ( G_{l+} |0\rangle \langle 0| G_{l+}^\dag \otimes |0\rangle\langle 0| \Big ) D^\dag \Big ] - \tr \Big [ E_0 D \Big ( G_{l-} |0\rangle\langle 0| G_{l-}^\dag \otimes |0\rangle\langle 0| \Big ) D^\dag \Big ] \Big \},
\end{split}
\label{e:grads_g}
\ee
where
\be
G_{l\pm} = G_L \cdots G_{l+1} G_{l}(\theta_{l}\pm \pi / 2) G_{l-1} \cdots G_{1}.
\ee
Note that $G_{l\pm}$ can be interpreted as two new circuits, each one differing from $G$ by an offset of $\pm \frac{\pi}{2}$ to parameter $\theta_l$. Hence, for each parameter $l$, we are required to execute the circuit compositions $D G_{l+}$ and $D G_{l-}$ on initial state ${|0\rangle^{\otimes n+1}}$ and measure the ancilla qubit. Because these auxiliary circuits have depth similar to that of the original circuit, estimation of the gradient is efficient. Interestingly, up to a scale factor of $\frac{\pi}{2}$, the analytical gradient is equal to the central finite difference approximation carried out at $\pi$.

Similarly, the analytical partial derivatives for the discriminator read
\be
\begin{split}
\frac{\partial V}{ \partial \phi_k} =& \frac{P(t)}{2} \Big \{ \tr \Big [ E_0 D_{k+} \Big ( |\psi_t\rangle\langle \psi_t| \otimes |0\rangle\langle 0| \Big ) D_{k+}^\dag \Big ] - \tr \Big[ E_0 D_{k-} \Big ( |\psi_t\rangle\langle \psi_t| \otimes |0\rangle\langle 0| \Big ) D_{k-}^\dag \Big ] \Big \} - \\
& \frac{P(g)}{2} \Big \{ \tr \Big [ E_0 D_{k+} \Big ( G |0\rangle\langle 0| G^\dag \otimes |0\rangle\langle 0| \Big ) D_{k+}^\dag \Big ] - \tr \Big[ E_0 D_{k-} \Big ( G |0\rangle\langle 0| G^\dag \otimes |0\rangle\langle 0| \Big ) D_{k-}^\dag \Big ] \Big \},
\end{split}
\label{e:grads_d}
\ee
where
\be
D_{k\pm} = D_K \cdots D_{k+1} D_{k}(\phi_{k}\pm \pi/2) D_{k-1} \cdots D_{1}.
\ee
In this case, for each parameter $k$ we are required to execute four auxiliary circuit compositions: $D_{k+}$ and $D_{k-}$ on target state $|\psi_t\rangle \otimes |0\rangle$, while $D_{k+} G$ and $D_{k-} G$ on initial state ${|0\rangle^{\otimes n+1}}$.

Finally, all parameters are updated by gradient descent/ascent
\be
\begin{split}
&\theta_l^{(t+1)} = \theta_l^{(t)} - \epsilon \frac{\partial V}{ \partial \theta_l} \Bigr|_{\footnotesize \btheta=\btheta^{(t)}, \bphi=\bphi^{(t)}}\\
&\phi_k^{(t+1)} = \phi_k^{(t)} + \eta \frac{\partial V}{ \partial \phi_k} \Bigr|_{\footnotesize \btheta=\btheta^{(t+1)}, \bphi=\bphi^{(t)}} ,
\end{split}
\label{e:updates}
\ee
where $\epsilon$ and $\eta$ are hyperparameters determining the step sizes. Here we rely on the fine-tuning of these, as opposed to Newton's method which makes use of the Hessian matrix to determine step sizes for all parameters. Other researchers~\cite{dallaire2018quantum} designed circuits to estimate the analytical gradient and the Hessian matrix. Such approach requires the ability to execute complex controlled operations and is expected to require error correction. Our approach and others'~\cite{mitarai2018quantum,liu2018differentiable} require much simpler circuits, which is desirable for implementation on NISQ computers.

As we discuss next, accelerated gradient techniques developed by the deep learning community can further improve our method. 

\subsection{Optimization by resilient backpropagation}

If we could minimize the trace distance in Eq.~\ref{e:tracedist} directly over the set of density matrices, then the problem would be convex~\cite{nielsen2011quantum}. However, in this paper we deal with a potentially non-convex problem due to the optimization of exponentiated parameters and hence the introduction of sine and cosine functions. 

A recent paper~\cite{mcclean2018barren} suggested that the error surface of circuit learning problems is challenging for gradient-based methods due to the existence of barren plateaus. In particular, the region where the gradient is close to zero does not correspond to local minima of interest, but rather to an exponentially large plateau of states that have exponentially small deviations in the objective value from that of the totally mixed state. While the derivation of the above statement is for a class of random circuits, in practice we prefer to deal with highly structured circuits~\cite{grant2018hierarchical,chen2018universal}. Moreover, here we argue that the existence of plateaus does not necessarily pose a problem for the learning of quantum circuits, provided that the sign of the gradient can be resolved. To validate this claim we refer to the classical literature and argue that similar problems have traditionally occurred also in classical neural network training and allow for efficient solutions.

Typical gradient-based methods update the parameters with steps of the form
\begin{equation}
w_{i}^{(t+1)} = w_{i}^{(t)} - \epsilon \frac{\partial}{\partial w_{i}} E^{(t)} ,
\end{equation}
where $w_{i}^{(t)}$ is the $i$-th parameter at time $t$, $\epsilon$ is the step size, $E^{(t)}$ is the error function to be minimized and its superscript indicates evaluation at $w=w^{(t)}$. If the step size is too small, the derivatives are also scaled to be too small resulting in a long time to convergence. If the step size is too large, this can lead to oscillatory behavior of the updates or even to divergence. One of the early approaches to counter this behavior was the introduction of a momentum term, which takes into account the previous steps when calculating the current update. 
The gradient descent with momentum (GDM) reads
\be
\label{e:gdm}
\begin{split}
&\Delta_i^{(t)} = - \epsilon \frac{\partial}{\partial w_{i}}E^{(t)} + \mu \Delta_i^{(t-1)} \\
&w_i^{(t+1)} = w_i^{(t)} + \Delta_i^{(t)} ,
\end{split}
\ee
where $\mu$ is a momentum hyperparameter. Momentum methods produce some resilience to plateaus in the error surface, but they lose this resilience when the plateaus are characterized by having very small or zero gradient.

A family of optimizers known as resilient backpropagation algorithms (Rprop)~\cite{riedmiller1993direct} is particularly well suited for problems where the error surface is characterized by large plateaus with small gradient. Rprop algorithms adapt the step size for each parameter based on the agreement between the sign of its current and previous partial derivatives. If the signs of the two derivatives agree, then the step size for that parameter is increased multiplicatively. This allows the optimizer to traverse large areas of small gradient with an increasingly high speed. If the signs disagree, it means that the last update for that parameter was large enough to jump over a local minima. To fix this, the parameter is reverted to its previous value and the step size is decreased multiplicatively. Rprop is therefore resilient to gradients with very small magnitude as long as the sign of the partial derivatives can be determined.

We use a variant known as iRprop$^{-}$~\cite{igel2000improving} which does not revert a parameter to its previous values when the signs of the partial derivatives disagree. Instead, it sets the current partial derivative to zero so that the parameter is not updated, but its step size is still reduced. The hyperparameters and pseudocode for iRprop$^{-}$ are described in Algorithm~\ref{alg:iRprop}.

\begin{algorithm}[H]
\caption{iRprop$^{-}$~\cite{igel2000improving}}
\begin{algorithmic}[1]
\Require {error function $E$, initial parameters $w_i^{(0)}$, initial step size $\Delta_{init}$, minimum allowed step size $\Delta_{min}$, maximum allowed step size $\Delta_{max}$, step size decrease factor $\eta^{-}$, and step size increase factor $\eta^{+}$}
\begin{spacing}{1.7}
\Ensure {$\Delta_i^{(-1)} := \Delta_{init}$ and $\frac{\partial}{\partial w_{i}}E^{(-1)} := 0$ for all $i$}
\Repeat
    \For {\textbf{each }$i$}
        \If{$\frac{\partial}{\partial w_{i}}E^{(t-1)} \frac{\partial}{\partial w_{i}}E^{(t)} > 0$}
            \State $\Delta_{i}^{(t)} := \min \{ \eta^{+} \Delta_{i}^{(t-1)}, \enskip \Delta_{max} \}$
        \ElsIf{$\frac{\partial}{\partial w_{i}}E^{(t-1)} \frac{\partial} {\partial w_{i}}E^{(t)} < 0$}
            \State $\Delta_{i}^{(t)} := \max \{ \eta^{-} \Delta_{i}^{(t-1)}, \enskip \Delta_{min} \}$
            \State $\frac{\partial}{\partial w_{i}}E^{(t)} := 0$
        \Else
            \State $\Delta_{i}^{(t)} := \Delta_{i}^{(t-1)}$
        \EndIf
        \State ${w_{i}}^{(t+1)} := {w_{i}}^{(t)} - \sgn(\frac{\partial}{\partial w_{i}}E^{(t)}) \Delta_{i}^{(t)}$
    \EndFor
\Until {convergence}
\end{spacing}
\vspace*{-.2cm}
\end{algorithmic}
\label{alg:iRprop}
\end{algorithm}

\clearpage

Despite the resilience of Rprop, if the magnitude of the gradient in a given direction is so small that the sign cannot be determined, then the algorithm will not take a step in that direction. Furthermore, the noise coming from the finite number of samples could cause the sign to flip at each iteration. This would quickly make the step size very small and the optimizer could get stuck on a barren plateau.

One possible modification is an explorative version of Rprop that explores areas with zero or very small gradient at the beginning of training, but still converges at the end of training. First, any zero or small gradient at the very beginning of training could be replaced by a positive gradient to ensure an initial direction is always defined. Second, one could use large step size factors and decrease them during training to allow for convergence to a minima. Finally, an explorative Rprop could remember the sign of the last suitably large gradient and take a step in that direction whenever the current gradient is zero. This way, when the optimizer encounters a plateau, it would traverse the plateau from the same direction it entered. We leave investigation of an explorative Rprop algorithm to future work.

\subsection{Heuristic for the stopping criterion}\label{s:stop_criterion}

Evaluating the performance of generative models is often intractable and can be done only via application-dependent heuristics~\cite{theis2016note,salimans2016improved}. This is also the case for our model as the value function in Eq.~\eqref{e:value_function} does not provide information about the generator's performance, unless the discriminator is optimal. Unfortunately, we do not always have access to an optimal discriminator (more on this in Section~\ref{s:error_corrected}). We now describe an efficient method that can be used to assess the learning in the quantum setting. In turn, this can be used to define a stopping criterion for the adversarial game.

We begin recalling that the discriminator makes use of projective measurements on an ancilla register $\mathcal{A}$ to effectively implement a POVM. Should the ancilla register be maximally entangled with the main register $\mathcal{M}$, its reduced density matrix would correspond to that of a maximally mixed state. Performing projective measurements on the maximally mixed state would then result in uniform random outcomes and decisions.

Ideally, the discriminator would encode all relevant information in the ancilla register and then remove all its correlations with the main register, obtaining a product state ${ \rho_d = \rho_d^\mathcal{M} \otimes \rho_d^\mathcal{A} }$. Hereby we use subscript $d$ to indicate the state output by the discriminator circuit. This scenario is similar in spirit to the uncomputation technique used in many quantum algorithms~\cite{bennett1997strengths}.

The bipartite entanglement entropy (BEE) is a measure that can be used to quantify how much entanglement there is between two partitions
\be
S(\rho_d^\mathcal{A}) = - \tr [ \rho_d^\mathcal{A} \ln \rho_d^\mathcal{A} ] = - \tr [ \rho_d^\mathcal{M} \ln \rho_d^\mathcal{M} ] = S(\rho_d^\mathcal{M}) ,
\ee
where $\rho_d^\mathcal{A} = \tr_\mathcal{M} [\rho_d]$ and $\rho_d^\mathcal{M} = \tr_\mathcal{A} [\rho_d]$ are reduced density matrices obtained by tracing out one of the partitions, i.e., by ignoring one of the registers. The BEE is intractable in general, but here we can exploit its symmetry and compute it on the smallest partition, i.e., the ancilla register $\mathcal{A}$. Because this register consists of a single qubit, BEE reduces to 
\be\label{e:bee}
S(\rho_d^\mathcal{A})=-\frac{1+\|\boldsymbol{r}\|}{2} \ln \Big ( \frac{1+\|\boldsymbol{r}\|}{2} \Big ) -\frac{1-\|\boldsymbol{r}\|}{2} \ln \Big ( \frac{1-\|\boldsymbol{r}\|}{2} \Big ) ,
\ee
where $\boldsymbol{r} \in {\rm I\!R}^3$ is the Bloch vector such that $\rho_d^\mathcal{A} = \frac{1}{2}(I + \boldsymbol{\sigma} \cdot \boldsymbol{r})$, $\|\boldsymbol{r}\| \leq 1$, and $\boldsymbol{\sigma} = (\sigma_x, \sigma_y, \sigma_z)$. The three components of the Bloch vector can be estimated using tomography techniques for a single qubit, for which we refer to the excellent review in Ref.~\cite{schmied2016quantum}.

There exist a wide range of methods that can be used depending on the desired accuracy, the prior knowledge, and the available computational resources. In this work we consider the scaled direct inversion (SDI)~\cite{schmied2016quantum} method, where each entry of the Bloch vector is estimated independently by measuring the corresponding Pauli operator. This is motivated by the fact that $\langle \sigma_i \rangle = \tr [\sigma_i \rho_d^\mathcal{A}] = \boldsymbol{e}_i \boldsymbol{r} $ where $\boldsymbol{e}_i$ is the Cartesian unit vector in the $i$ direction and $i \in \{ x, y, z \}$. These measurements can be done in all existing gate-based quantum computers we are aware of by applying a suitable rotation followed by a measurement in the computational basis. 

We can write a temporary Bloch vector $\widehat{\boldsymbol{r}}_0 = ( \widehat{\langle\sigma_x\rangle}, \widehat{\langle\sigma_y\rangle}, \widehat{\langle\sigma_z\rangle})$ where all expectations are estimated from samples. Due to finite sampling error, there is non-zero probability that the vector lies outside the unite sphere, although inside the unit cube. These cases correspond to non-physical states and SDI corrects them by finding the valid state with minimum distance over all Schatten p-distances. It turns out, this is simply the rescaled vector~\cite{schmied2016quantum}
\be
\widehat{\boldsymbol{r}} =
\begin{cases}
\widehat{\boldsymbol{r}}_0 \quad &\text{if} \quad \|\widehat{\boldsymbol{r}}_0\| \leq 1 \\
\widehat{\boldsymbol{r}}_0 / \|\widehat{\boldsymbol{r}}_0\|  &\text{otherwise.} \\
\end{cases}
\ee
The procedure discussed so far allows us to efficiently estimate the BEE in Eq.~\eqref{e:bee}. Equipped with this information, we can now design an heuristic for the stopping criterion. 

The reasoning is as follows. Provided that the discriminator circuit has enough connectivity, random initialization of its parameters will likely generate entanglement between main and ancilla registers. In other words, $S(\rho_d^\mathcal{A})$ is expected to be large at the beginning. As the learning algorithm iterates, the discriminator gets more accurate at distinguishing states. As discussed above, this requires the ancilla qubit to depart from the totally mixed state and $S(\rho_d^\mathcal{A})$ to decrease. This is when the learning signal for the generator is stronger, allowing the generated state to get closer to the target. As the two become less and less distinguishable with enough iterations, the discriminator needs to increase correlations between ancilla's bases and relevant factors in the main register. That is, we expect to observe an increase of entanglement between the two registers, hence an increase in $S(\rho_d^\mathcal{A})$. The performance of the discriminator would then saturate as $S(\rho_d^\mathcal{A})$ converges to its upper bound of $\ln(2)$. We propose to detect this convergence and use it as a stopping criterion. In the Section~\ref{s:results} we analyze the behavior of BEE via numerical simulations.

\subsection{Long-term implementation on universal quantum computers}\label{s:error_corrected}

Let us briefly recall the adversarial circuit learning task. We have two circuits, the generator and the discriminator, and a target state. The target state $\rho_t$ is prepared with probability $P(t)$, while the generated state $\rho_g$ is prepared with probability $P(g)$. The discriminator has to successfully distinguish each state or, in other words, he must find the measurement that minimizes the probability of labelling error.

As described earlier, Helstrom~\cite{helstrom1969quantum} observed that the optimal POVM that distinguishes two states has the following particular form; Let $E_0$ and $E_1$ be the POVM elements attaining the minimum in $P^*_{err} = \min_{\{E_0, E_1\}} \tr[E_1\rho_t]P(t) + \tr[E_0\rho_g]P(g)$, then both elements are diagonal in a basis that also diagonalizes the Hermitian operator
\be
\Gamma = P(t) \rho_t - P(g) \rho_g .
\ee
As pointed out in Ref.~\cite{fuchs1996distinguishability}, in this basis one can construct $E_0$ by specifying its diagonal elements $\lambda_j$ according to the rule
\be
\begin{split}
&\lambda_j = 1 \quad \text{when} \quad \gamma_j < 0 \\
&\lambda_j = 0 \quad \text{when} \quad \gamma_j \geq 0 ,
\end{split}
\ee
where $\gamma_j$ are the diagonal elements of $\Gamma$. The operator $E_1$ is then obtained via the relationship $I-E_0$. Hence we can construct the optimal measurement operator if we have access to the operator $\Gamma$, and provided that we can \mbox{diagonalize it.}

Using the above insight, with ${ \rho_t = | \psi_t \rangle \langle \psi_t| }$ and ${ \rho_g = | \psi_g \rangle \langle \psi_g | }$, we can observe that ${ \tr[\Gamma \rho_g] = P(t) | \langle \psi_g | \psi_t \rangle |^2 - P(g) }$ and 
${ \tr[\Gamma \rho_t] = P(t) - P(g) | \langle \psi_g | \psi_t \rangle |^2 }$.
Under the assumption of equal prior probabilities of $1/2$, the above is minimized for a maximum overlap of the two states. Since the prior probabilities are hyperparameters, we can set them to $1/2$ and use the swap test~\cite{buhrman2001quantum} to compute the overlap. This procedure effectively implements an optimal discriminator and provides a strong learning signal to the generator. 

Note, however, that the swap test bears several disadvantages. In order to perform the swap test, we need to access both $\rho_t$ and $\rho_g$ simultaneously. This also requires the use of two registers for a total $2n+1$ qubits, which is significantly more than the $n+1$ qubits required in the near-term approach. Finally, the swap test requires the ability to perform non-trivial controlled gates and error correction.

A potential solution is to find an efficient low-depth circuit implementing the swap test. In Ref.~\cite{cincio2018learning} the authors implemented such via a variationally trained circuit. As pointed out in their work, this requires (a) an order of $2^{2n}$ training examples for states of $n$ qubits, and (b) each training example be given by the actual overlap between two states, requiring a circuit which gives the answer to the problem we are trying to solve. We hence hold the belief that this approach is not suitable for our task. However, other approaches for finding a low-depth circuit for computing the swap test might well be possible.

One could alternatively consider the possibility of implementing a discriminator via distance measurements based on random projections, i.e., Johnson-Lindenstrauss transformations~\cite{dasgupta2003elementary}. This would require a reduced amount of resources and could be adapted for the adversarial learning task. As an example, we could apply a quantum channel to coherently reduce the dimensionality of the input state and then apply the state discrimination procedure in the lower dimensional space. However, in Ref.~\cite{harrow2011limitations} the authors proved that such an operation cannot be performed by a quantum channel. One way to think about this is that the Johnson-Lindenstrauss transformation is a projection onto a small random subspace and therefore a projective measurement. As the subspace is exponentially smaller than the initial Hilbert space, the probability that this projection preserves the distances is very small.

\section{Results}\label{s:results}

\begin{figure*}[t]
 \raisebox{-0.5\height}{\includegraphics[width=.64\textwidth]{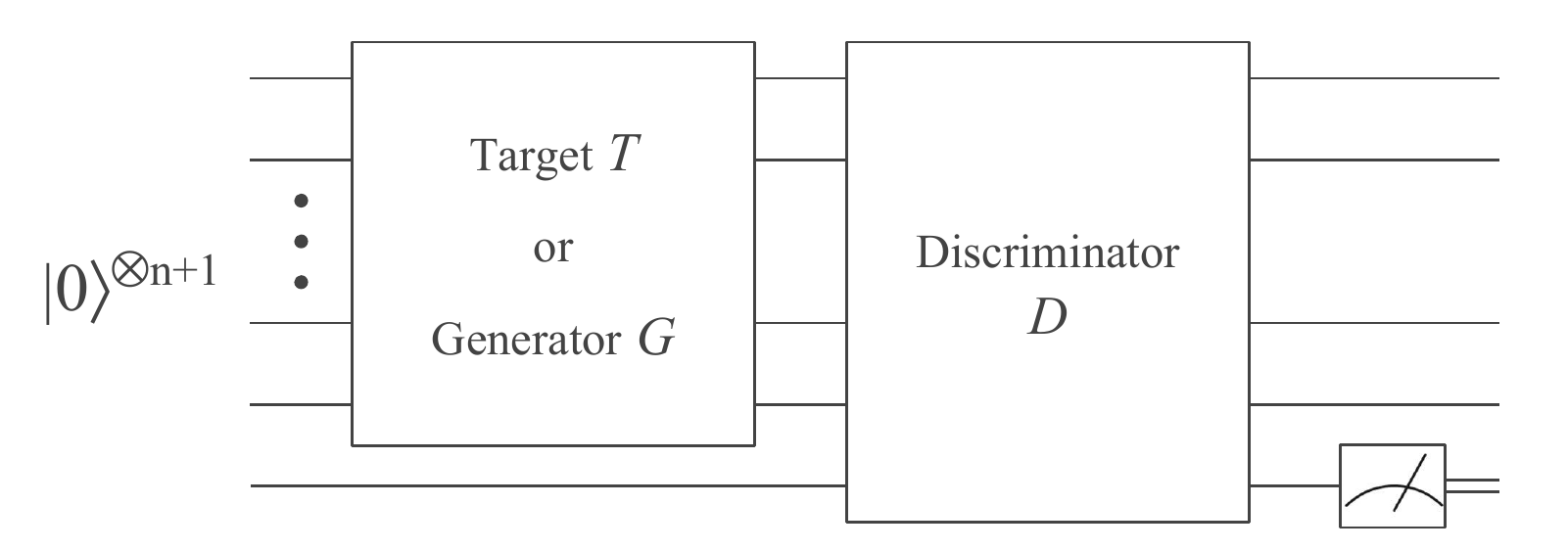}}
\hspace*{35px}
 \raisebox{-0.5\height}{\includegraphics[width=.23\textwidth]{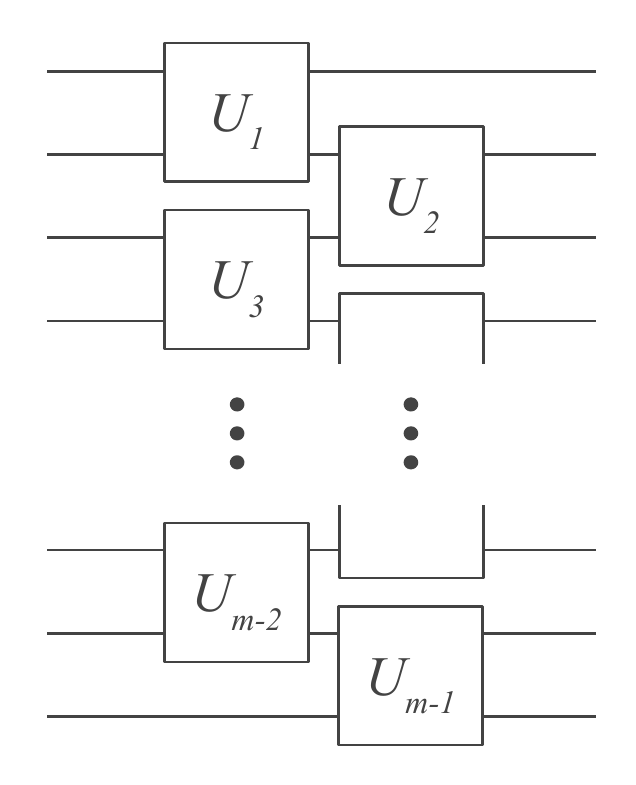}}
\caption{Left panel: Representation of the adversarial quantum circuits. In our simulations the target state is prepared by a random circuit $T$. The generator circuit $G$ learns to approximate the target. The discriminator circuit $D$ takes in input unknown $n$-qubit states and learns to label them as `target' or `generated'. This is done via the binary outcome of a projective measurement on a single ancilla qubit. Neither the generator nor the discriminator are allowed to `see' the inner workings of $T$ at any time. Hence, the learning signal for the generator comes solely from the probability of error of the discriminator. Right panel: Layout used as a building block for all the circuits. For an $m$-qubit circuit the layer has $m-1$ general two-qubit unitaries. General two-qubit unitaries of this kind can be efficiently implemented with three CNOT gates and $15$ parametrized single-qubit rotations as in Ref.~\cite{shende2004minimal}.}
\label{f:setting}
\end{figure*}

We show that adversarial quantum circuit learning can be used to approximate entangled target states. In realistic scenarios, the target state would come from an external channel and would be loaded in the quantum computer's register with no significant overhead. For the simulations we mock this scenario using circuits to prepare the target states. That is, we have $\rho_t = T |0\rangle\langle 0| T^\dag$ where $T$ is an unknown circuit. We setup a generator circuit $G$ and a discriminator circuit $D$, and the composition of these circuits is shown in Fig.~\ref{f:setting}, left panel. We shall stress that neither the generator nor the discriminator are allowed to `see' the inner workings of $T$ at any time.

We are interested in studying the performance of the algorithm as we change the complexity of the circuits. The complexity of our circuits is determined by the number of layers of gates. We denote such a number as $c(\cdot)$ so that, for example, a generator circuit $G$ made of $2$ layers has complexity $c(G)=2$. Figure~\ref{f:setting}, right panel, shows the layer that we used for our circuits. It has $m-1$ general two-qubit gates where $m$ is the number of qubits. Note that a general two-qubit gate can be efficiently implemented with three CNOT gates and $15$ parametrized single-qubit rotations as shown in Ref.~\cite{shende2004minimal}.

All parameters were initialized uniformly at random in $[-\pi,+\pi]$. We chose ${P(t)=P(g)=\frac{1}{2}}$ so that the discriminator is given target and generated states with equal probability. All expected values required to compute gradients were estimated from $100$ measurements on the ancilla qubit. Unless stated otherwise, optimization was performed using iRprop$^{-}$. We used an initial step size \mbox{$\Delta_{init}=1.5\pi \times 10^{-3}$}, a minimum allowed step size \mbox{$\Delta_{min} =\pi \times 10^{-6}$}, and a maximum allowed step size \mbox{$\Delta_{max} =6\pi \times 10^{-3}$}.

\begin{figure*}[t]
\includegraphics[width=.99\textwidth]{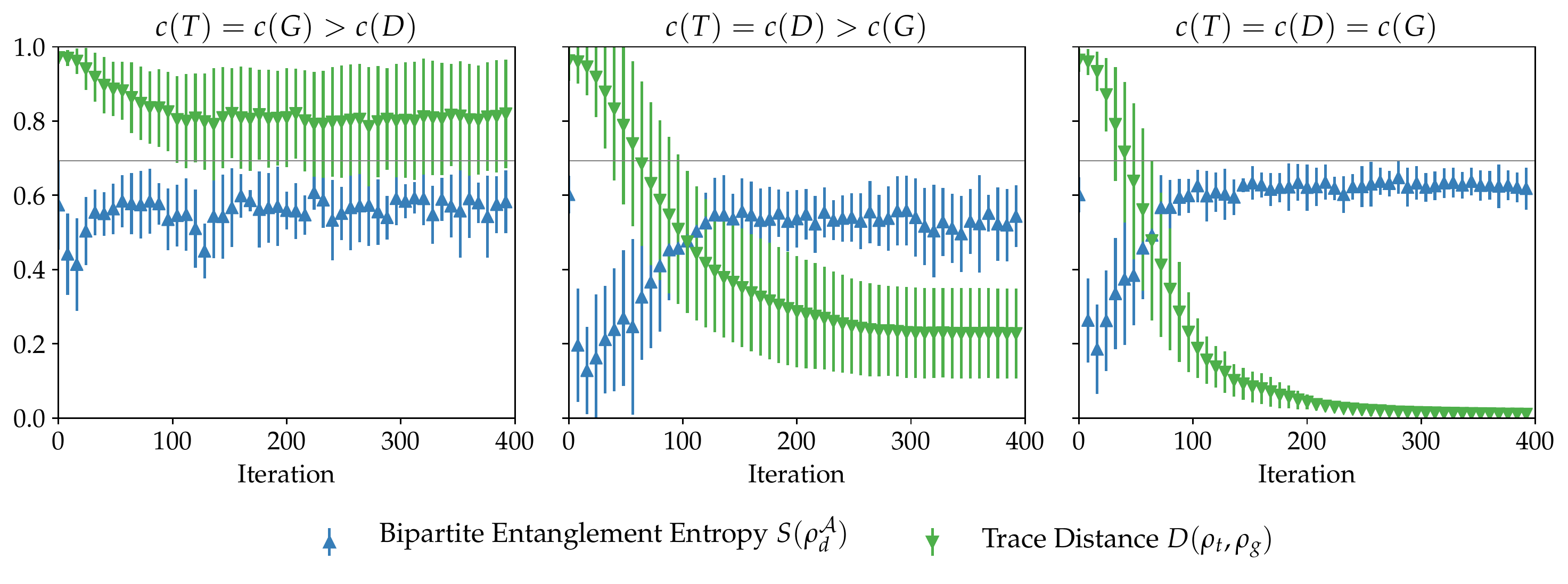}
\caption{Learning curves and stopping criterion for simulations on four-qubit target states. The performance is shown in terms of the trace distance between the target and generated states (green downward triangles), with zero indicating optimal approximation. All lines represent mean and one standard deviation computed on $10$ repetitions. Titles indicate the complexities of target $c(T)$, generator $c(G)$, and discriminator $c(D)$ circuits (see main text for details). In the left panel, the discriminator is too simple to provide a learning signal for the generator. In the central panel, the generator is simple, but it can still produce a meaningful approximation of the target state. In the right panel, all circuits are complex enough to learn an indistinguishable approximation of the target state. The trace distance cannot be computed in near-term implementations. The bipartite entanglement entropy (BEE) of the ancilla qubit (blue upward triangles) can be used as an efficient proxy to assess the learning progress. After the initial drop in BEE, the learning signal for the generator is strong and the trace distance decreases sharply. As learning progresses, the ancilla qubit gets closer to the mixed state where  $S(\rho_d^\mathcal{A}) = \ln (2) \approx 0.69$ (gray horizontal line). Detecting the convergence of BEE can be used as a stopping criterion for training.}
\label{f:bee}
\end{figure*}

Figure~\ref{f:bee} shows learning curves for simulations on four qubits. The green downward triangles represent mean and one standard deviation of the trace distance between target and generated state, computed on $10$ repetitions. In the left panel, the number of layers are ${ c(T)=c(G)=2 }$ and $c(D)=1$. We observe that the complexity of the discriminator is not sufficient to provide a learning signal for the generator, and the final approximation is indeed not satisfactory. In the central panel, ${ c(T)=c(D)=2 }$ and $c(G)=1$. The generator is less complex than the target state, but it manages to produce a meaningful approximation in average. In the right panel, ${ c(T)=c(G)=c(D)=2 }$. The complexity of all circuits is optimal, and the generator learns an indistinguishable approximation of the target state. 

The trace distance reported here could have been approximately computed using the swap test. However, since we assumed a near-term implementation, we cannot reliably execute the swap test. In Section~\ref{s:stop_criterion} we designed an efficient heuristic to keep track of learning and suggested to use it as a stopping criterion. To test the idea, we performed additional $100$ measurements on the ancilla qubit for each observable $\sigma_x$, $\sigma_y$, and $\sigma_z$. The outcomes were used to estimate the BEE using the SDI method. In Fig.~\ref{f:bee} the blue upwards triangles represent mean and one standard deviation of the BEE, computed on $10$ repetitions. The left panel shows that when the discriminator circuit is too shallow, BEE oscillates with no clear pattern. The central and right panels show that, when using a favorable setting, the initial BEE drops significantly towards zero. This is when the generator begins to learn the target state. Note that, as the algorithm iterates, the ancilla qubit tends towards the maximally mixed state where ${ S(\rho_d^\mathcal{A}) = \ln (2) \approx 0.69 }$ (gray horizontal line). In this regime, the discriminator predicts the labels with probability equal to the prior ${ P(t)=P(g)=\frac{1}{2} }$.

\begin{figure*}
\includegraphics[width=.325\textwidth]{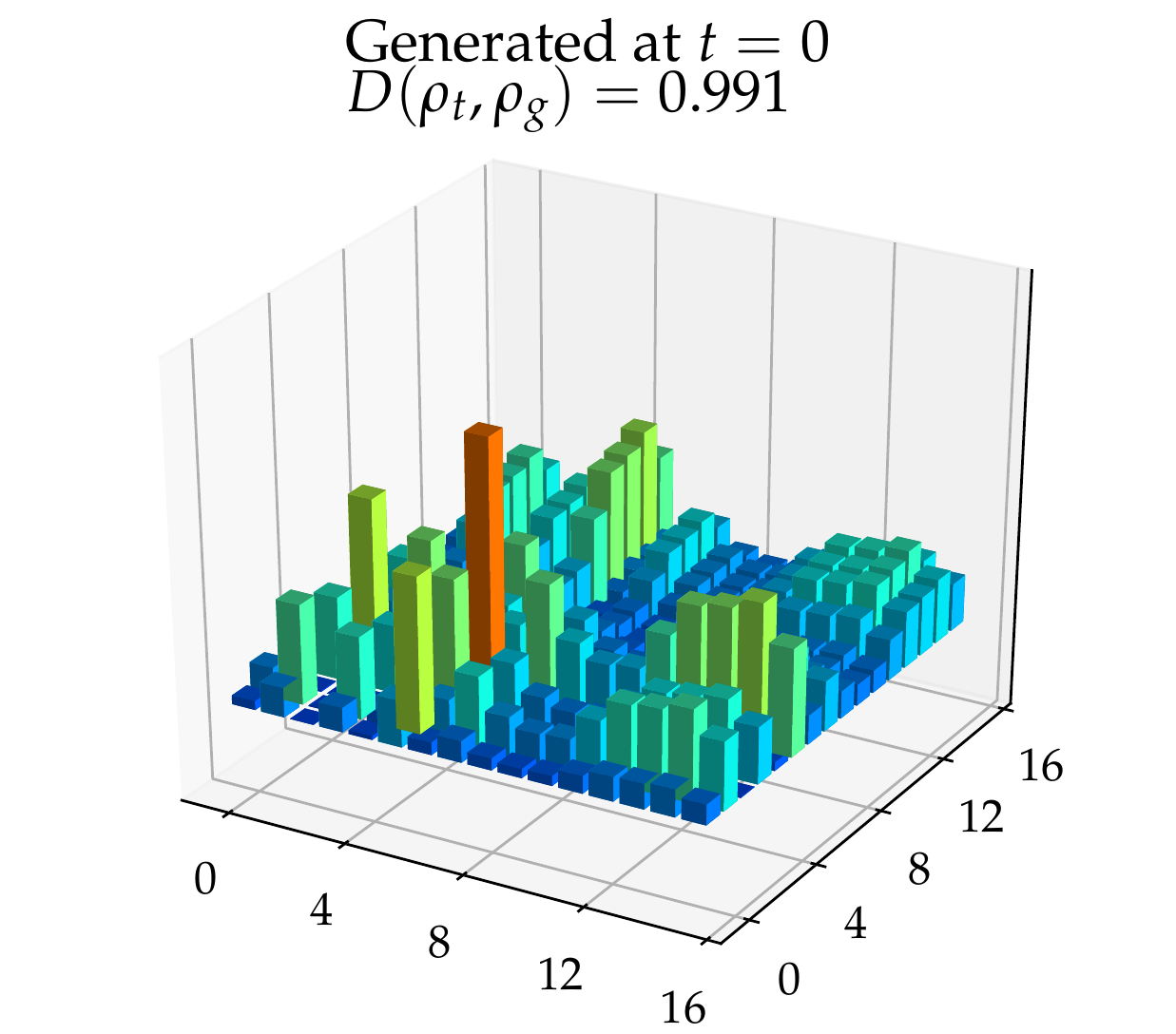}
\includegraphics[width=.325\textwidth]{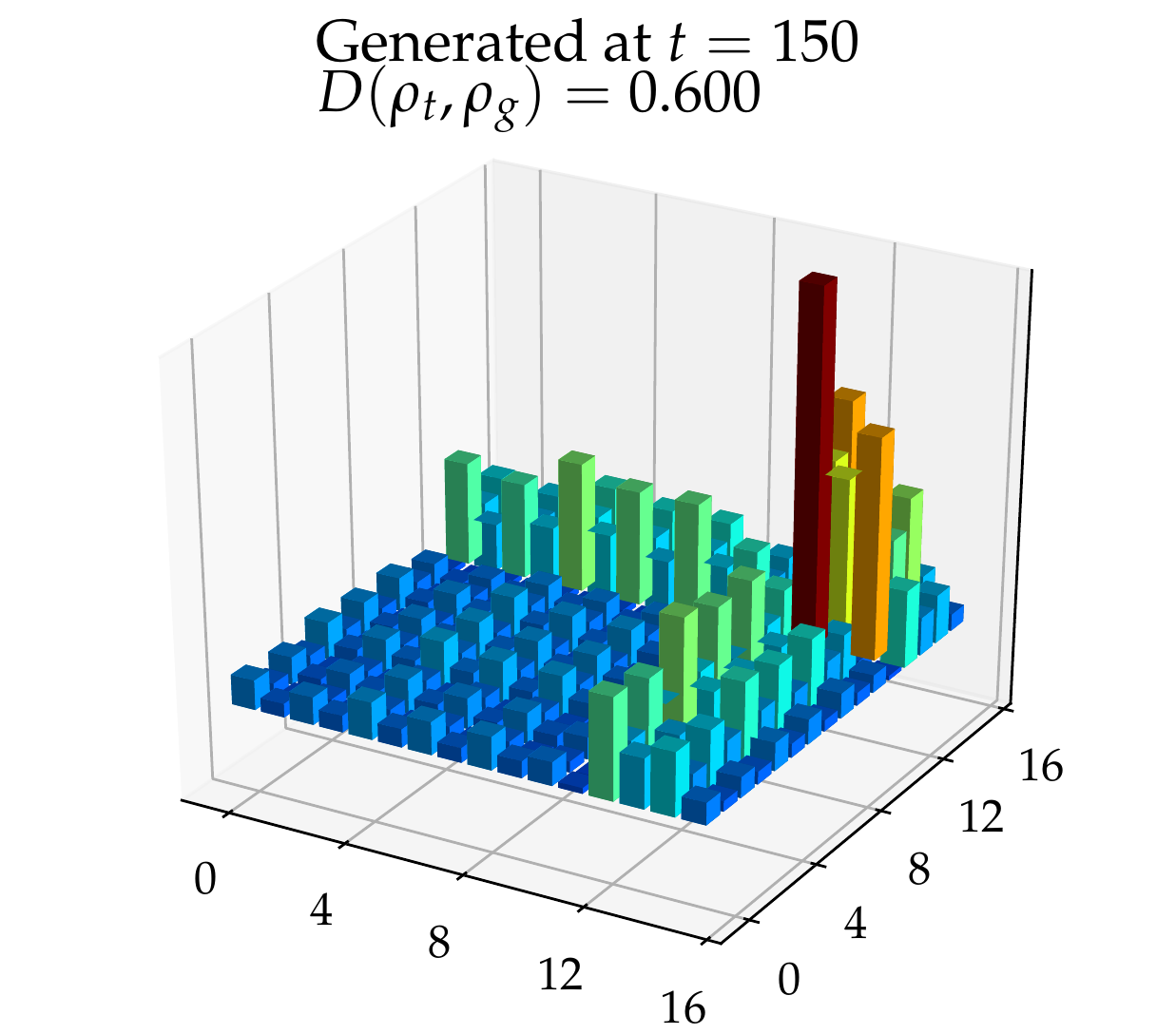}
\includegraphics[width=.325\textwidth]{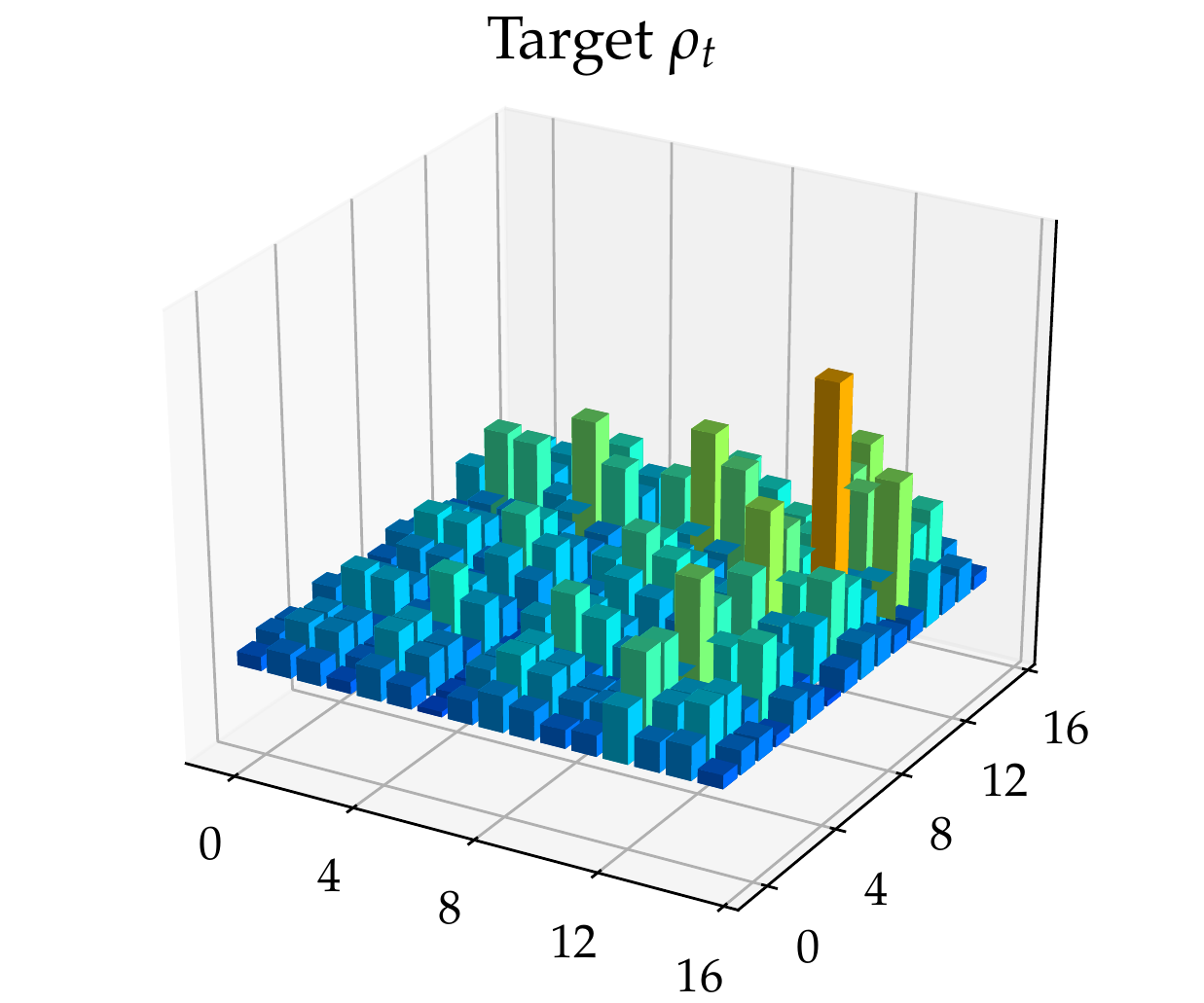}
\caption{Absolute value of tomographic reconstructions for a four-qubit target state. The target state is prepared by a random circuit of $c(T)=2$ layers (see main text for details), and the absolute value of its density matrix is shown in the right panel. The two players of the adversarial game are a generator with $c(G)=1$ and a discriminator with $c(D)=2$. The generator is too simple to learn the target exactly, but can still find a reasonable approximation. The initial generated state shown in the left panel is at trace distance $0.991$ from the target. Using our heuristic we stopped the adversarial learning at iteration $150$ where BEE converged. The final state, shown in the central panel, is at trace distance $0.6$ from the target. The generator managed to capture the main mode of the density matrix, that is, the sharp peak visible on the right.}
\label{f:tom_aprox}
\end{figure*}

\begin{figure*}
\includegraphics[width=.325\textwidth]{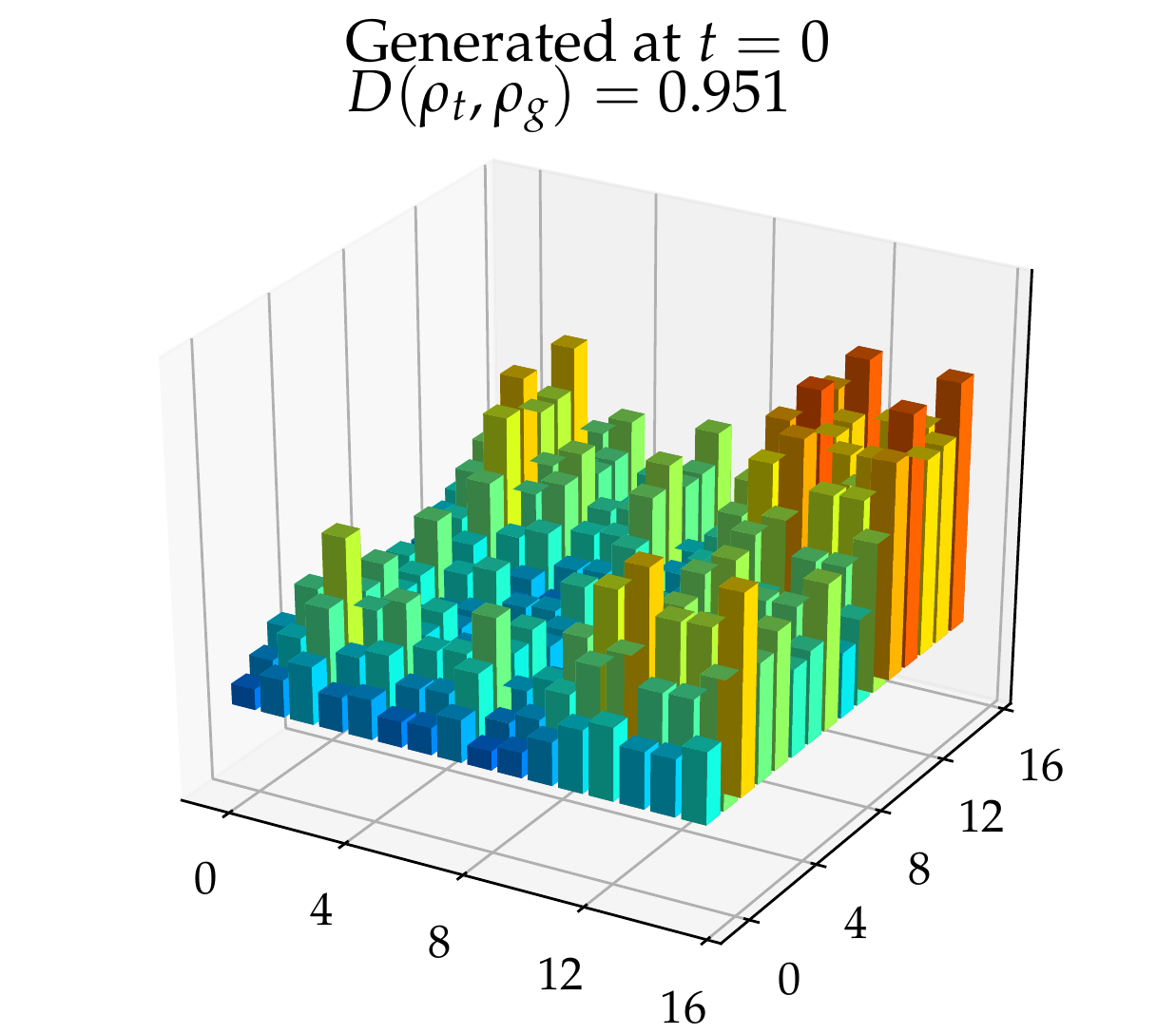}
\includegraphics[width=.325\textwidth]{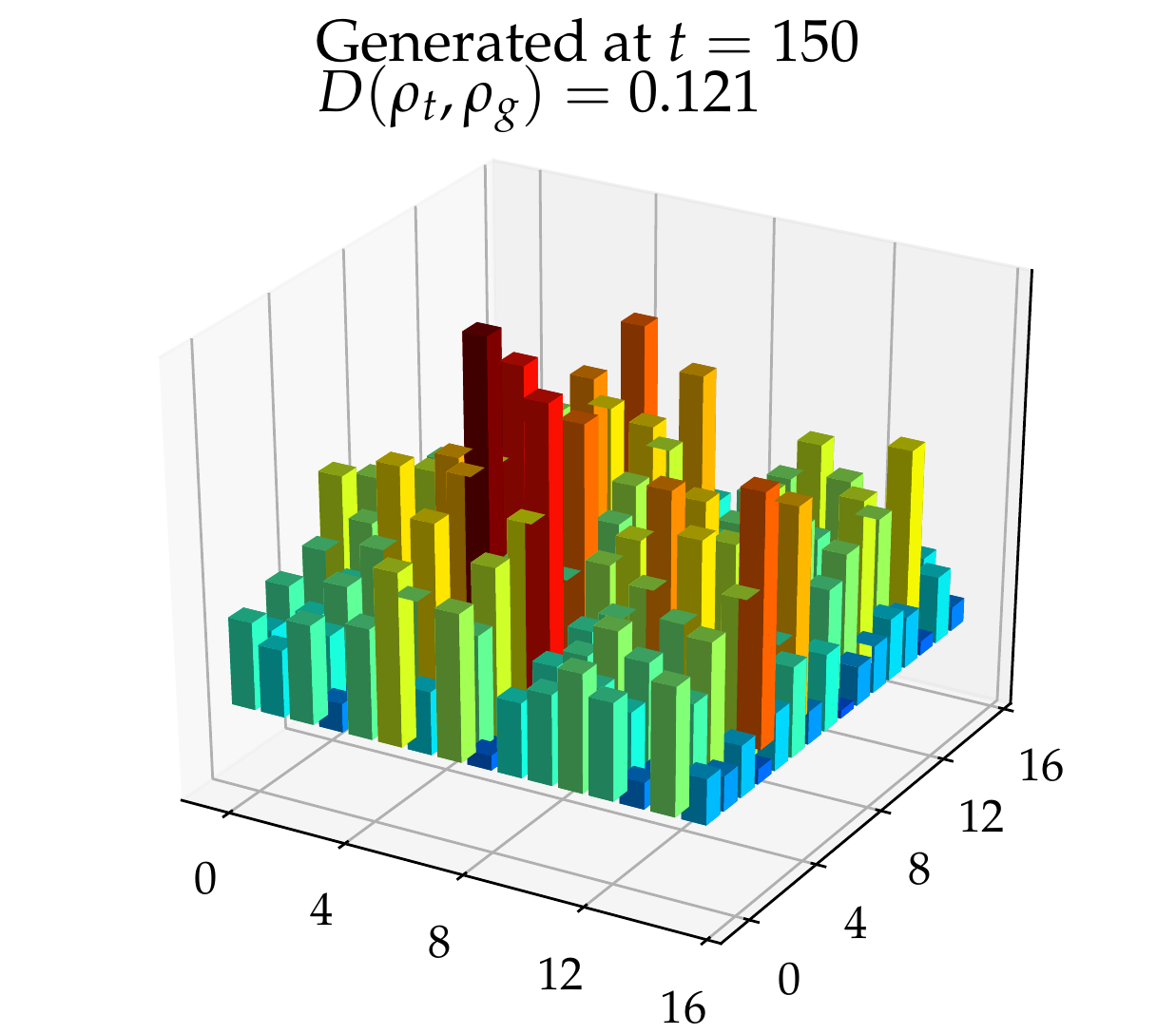}
\includegraphics[width=.325\textwidth]{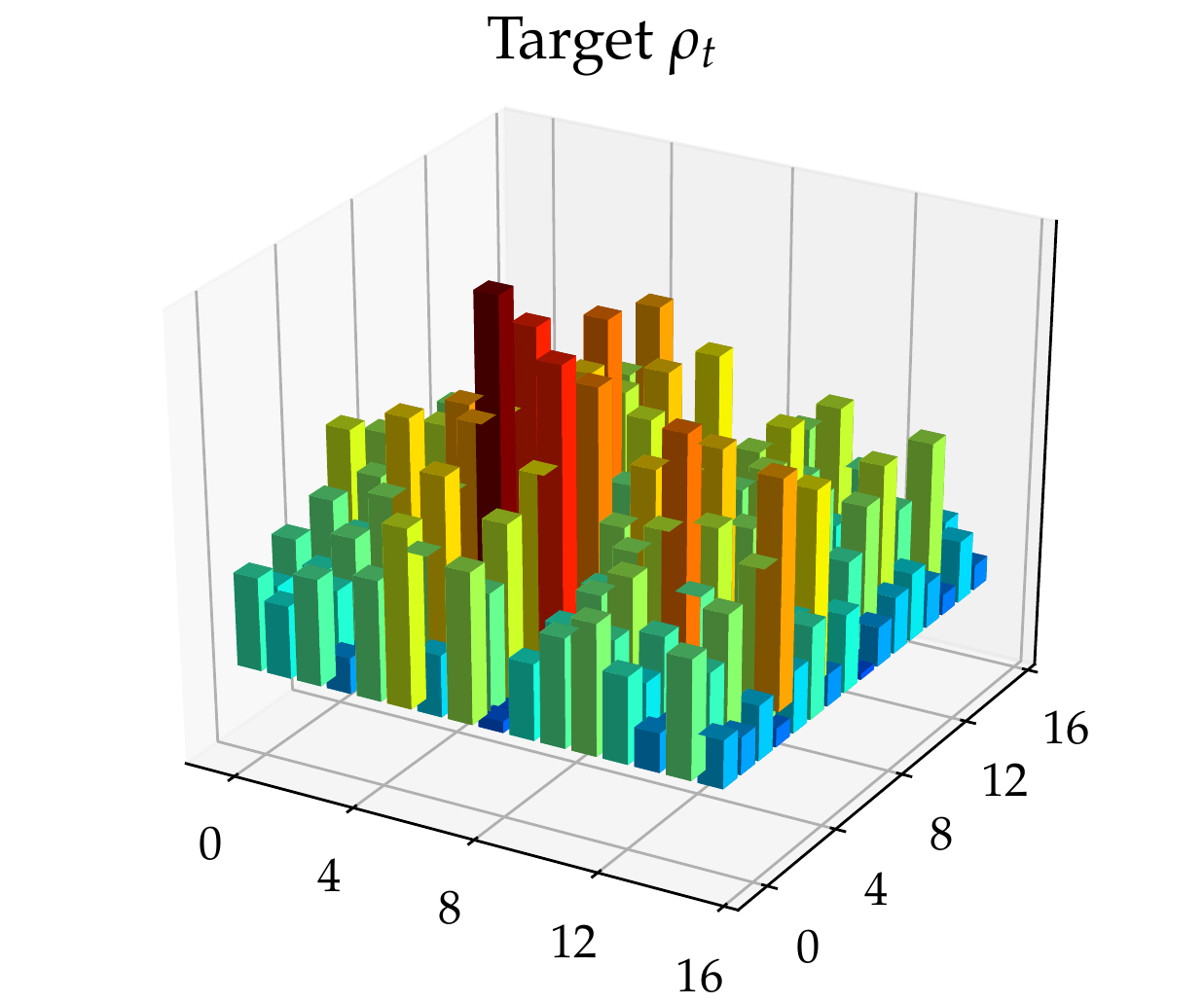}
\caption{Absolute values of tomographic reconstructions for a four-qubit target state. The setting is similar to that of Fig.~\ref{f:tom_aprox}, but this time the generator is a circuit with $c(G)=2$ layers, just like the random circuit that prepared the target. The randomly initialized generator produces the state shown in the left panel, which is at trace distance $0.951$ from the target. Using our heuristic we stopped the adversarial learning at iteration $150$ where BEE converged.  The final state, shown in the central panel, is at trace distance $0.121$ from the target. Visually, the target and final states are indistinguishable.}
\label{f:tom_good}
\end{figure*}

Detecting convergence of BEE can be used as a stopping criterion for training. For example, the central and right panels in Fig.~\ref{f:bee} show that BEE converged after approximately $150$ iterations. Stopping the simulation at that point we obtained excellent results in average. We now show tomographic reconstructions for two cases. First, we examine the case where the generator is under-parametrized. Figure~\ref{f:tom_aprox}, right panel, shows the absolute value of the entries of the density matrix for a four-qubit target state. The randomly initialized generator produced the state shown in the left panel which is at $0.991$ trace distance from the target. By stopping the adversarial algorithm after $150$ iterations, we generated the state shown in the central panel whose trace distance is $0.6$. The generator managed to capture the main mode of the density matrix, that is, the sharp peak visible on the right. Second, we examine the case where the generator is sufficiently parametrized. Figure~\ref{f:tom_good}, right panel, shows the absolute value of the entries of the density matrix for the target state. The generator initially produced the state shown in the left panel which is at trace distance $0.951$ from the target. By stopping the adversarial algorithm after $150$ iterations, we generated the state shown in the central panel whose trace distance is $0.121$. Visually, the target and final states are indistinguishable.

\begin{figure*}
\includegraphics[width=.33\textwidth]{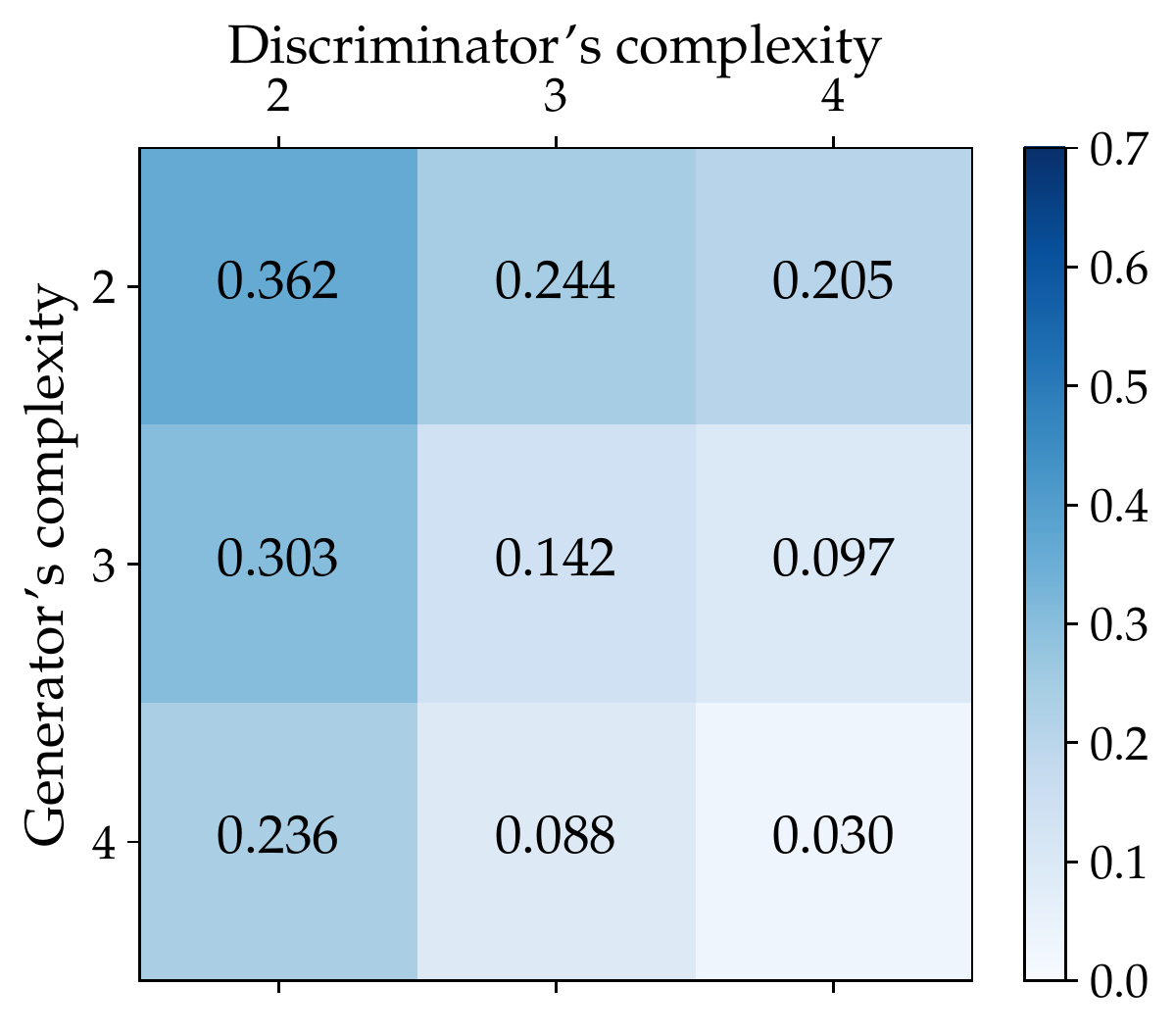}
\caption{Quality of the approximation against complexity of circuits for simulations on six-qubit target states. The heat-map shows mean trace distance of five repetitions of adversarial learning computed at iteration $600$. All standard deviations were $<0.1$ (not shown). The targets were produced by random circuits of $c(T)=3$ layers. Increasing the complexity of discriminator $c(D) \in \{2,3,4\}$ and generator $c(G) \in \{2,3,4\}$ resulted in better approximations to the target state in all cases.}
\label{f:complexity_study}
\end{figure*}

But how do the complexities of generator and discriminator affect the outcome? To verify this, we run the adversarial learning on six-qubit target states of $c(T)=3$ layers, and varied the number of layers of generator and discriminator. After $600$ training iterations, we computed the mean trace distance across five repetitions. As illustrated in Fig.~\ref{f:complexity_study}, increasing the complexity always resulted in a better approximation to the target state.

\begin{figure*}
\includegraphics[width=.5\textwidth]{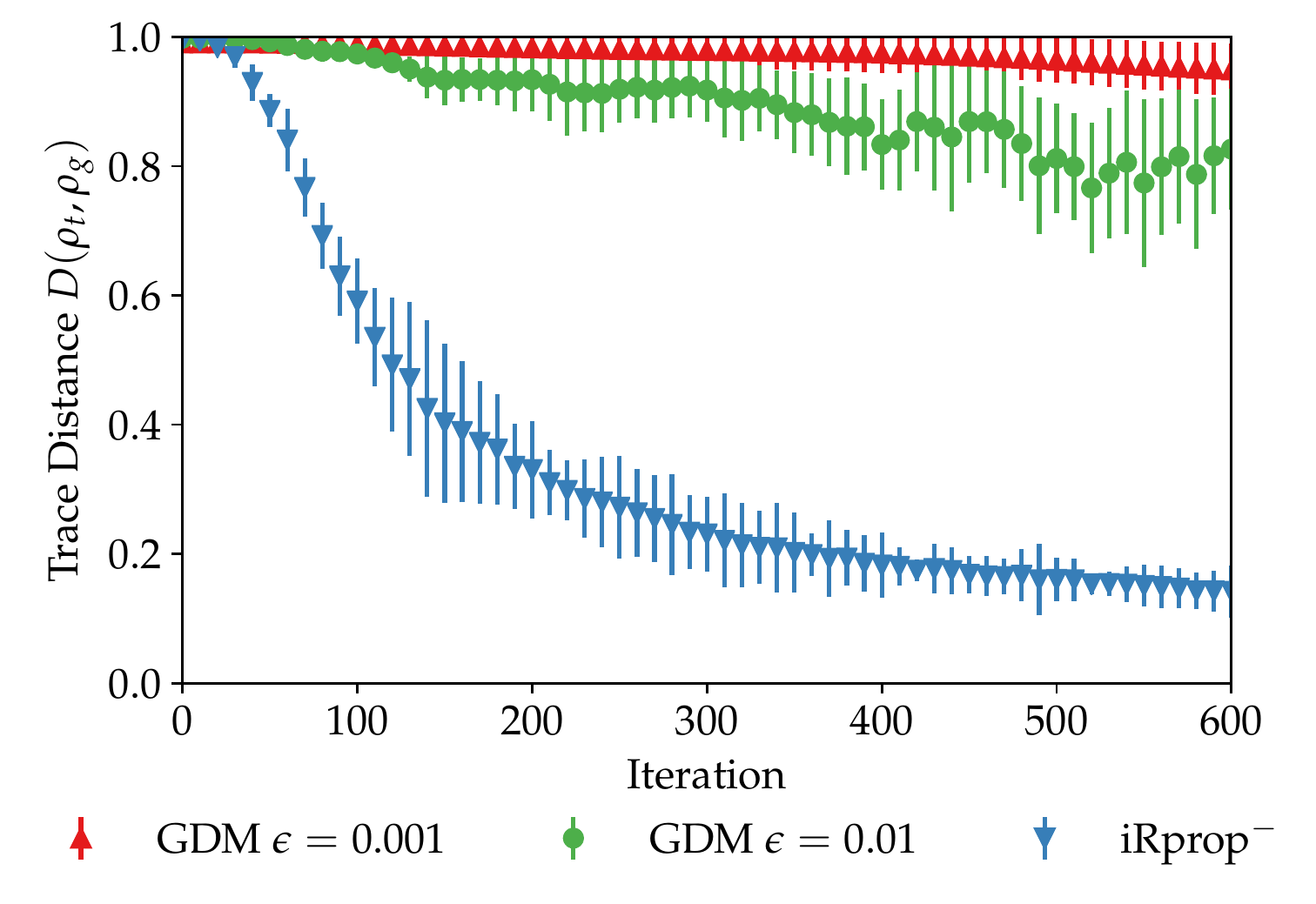}
\caption{Learning curves for different optimizers in simulations on six-qubit target states. The lines represent mean and one standard deviation of the trace distance computed on five repetitions. All circuits had the same number of layers, ${ c(T)=c(G)=c(D)=3 }$. iRprop$^{-}$ resulted in better performance than gradient descent with momentum (GDM) when using two different step sizes. Increasing the step size further in GDM resulted in unstable performance (not shown).}
\label{f:optimizer_study}
\end{figure*}

In our final test, we compared optimization algorithms on six-qubit target states. We ran GDM and iRprop$^{-}$ for $600$ iterations. Figure~\ref{f:optimizer_study} shows mean and one standard deviation across five repetitions. iRprop$^{-}$ (blue downward triangles) outperformed GDM both with step size $\epsilon=0.01$ (green circles) and $\epsilon=0.001$ (red upward triangles). This is because despite the small magnitude of the gradients when considering targets of six qubits, we were still able to estimate their sign and take relevant steps in the correct direction. This is a significant advantage of resilient backpropagation algorithms. 

\newpage

We now briefly discuss the advantages of our method compared to other quantum machine learning approaches for state approximation. These approaches require quantum resources that go far beyond those currently available. For example, the quantum principal component analysis~\cite{lloyd2014quantum} requires universal fault-tolerant hardware in order to implement the necessary SWAP operations. As another example, the quantum Boltzmann machine~\cite{amin2018quantum,kieferova2017tomography} requires the preparation of highly non-trivial thermal states. Moreover, those approaches provide limited control over the level of approximation. In contrast, the adversarial method proposed here is an heuristic scheme with fine control over the level of approximation; this is done by fixing the depth of the circuit, thereby limiting the complexity of the optimization problem. In this way, our method is expected to scale to large input dimensions, although this may require introducing an approximation error. As shown in Figs.~\ref{f:bee} and~\ref{f:complexity_study}, the error is an increasing function of the target's complexity, and a decreasing function of the generator's complexity. This feature allows the adversarial approach to be implemented with any available circuit depth on any NISQ device. A circuit-based demonstration of adversarial learning was given in Ref.~\cite{hu2019quantum} after our work. Clearly, a thorough numerical benchmark is needed to compare the scalability of different methods, which we leave for future work.

\section{Discussion and Conclusions}

In this work we proposed an adversarial algorithm and applied it to learn quantum circuits that can approximately generate and discriminate pure quantum states. We used information theoretic arguments to formalize the problem as a minimax game. The discriminator circuit maximizes the value function in order to better distinguish between the target and generated states. This can be thought of as learning to perform the Helstrom measurement~\cite{helstrom1969quantum}. In turn, the generator circuit minimizes the value function in order to deceive the discriminator. This can be thought of as minimizing the trace distance of the generated state to the target state. The desired outcome of this game is to obtain the best approximation to the target state for a given generator circuit layout.

We demonstrated how to perform such a minimax game in near-term quantum devices, i.e., NISQ computers~\cite{preskill2018quantum}, and we discussed long-term implementations on universal quantum computers. The near-term implementation has the advantage that it requires less qubits and avoids the swap test. The long-term implementation has the advantage that it can make use of the actual Helstrom measurement, with the potential of speeding up the learning process. 

Previous work on quantum circuit learning raised the concern of barren plateaus in the error surface~\cite{mcclean2018barren}. We showed numerically that a class of optimizers called resilient backpropagation~\cite{riedmiller1993direct} achieves high performance for the problem at hand, while gradient descent with momentum performs relatively poorly. These resilient optimizers require only the temporal behaviour of the sign of the gradient, and not the magnitude, to perform an update step. In our simulations of up to seven qubits we were able to correctly ascertain the sign of the gradient frequently enough for the optimizer to converge to a good solution. For regions of the error surface where the sign of the gradient cannot be reliably determined, we suggested an alternative optimization method that could traverse such regions. We will explore this idea in future work.

In general it is not clear how to assess the model quality in generative adversarial learning, nor how to find a stopping criterion for the optimization algorithm. For example, in the classical setting of computer vision, it is often the case that generated samples are visually evaluated by humans, i.e., the Turing test, or by a proxy artificial neural network, e.g., the Inception Score~\cite{salimans2016improved}. The quantum setting does not allow for these approaches in a straightforward manner. We therefore designed an efficient heuristic based on an estimate of the entanglement entropy of a single qubit, and numerically showed that convergence of this quantity indicates saturation of the adversarial algorithm. We therefore propose this approach as a stopping criterion for the optimization process. We conjecture that similar ideas could be used for regularization in quantum circuit learning for classification and regression.

We tested the quality of the approximations as a function of the complexity of the generator and discriminator circuits for simulations of up to seven qubits. Our results indicate that investing more resources in the generator and discriminator circuits leads to noticeable improvements. Indeed, an interesting avenue for future work is the study of circuit layouts, i.e., type of gates, and parameter initializations. If prior information about the target state is available, or can be efficiently extracted, we can encode it by using a suitable layout for the generator circuit. For example, in Ref.~\cite{liu2018differentiable} the authors use the Chow-Liu tree to displace CNOT gates such that they capture most of the mutual information among variables. Similarly, structured layouts could be used for the discriminator circuit such as hierarchical~\cite{grant2018hierarchical} and universal topologies~\cite{chen2018universal}. These choices could reduce the number of parameters to learn and simplify the error surface.

An adversarial learning framework capable of handling mixed states has been recently put forward~\cite{lloyd2014quantum,dallaire2018quantum}, but no implementation compatible with near-term computers was provided. In comparison, our framework works well for approximating pure target states and can find application in quantum state tomography on NISQ computers.

In this work we relied on the variational definition of Bayesian probability of error, which assumes the availability of a single copy of the quantum state to discriminate. By assuming the availability of multiple copies, which is in practice the case, one can derive more general adversarial games based on complex information theoretical quantities. These could be variational definitions of the quantum Chernoff bound~\cite{audenaert2007discriminating}, the Umegaki relative information, and other measures of distinguishability~\cite{fuchs1996distinguishability}.

\section{Acknowledgements}
The authors want to thank Ashley Montanaro for helpful discussions on random projections and for pointing out reference~\cite{harrow2011limitations}. M.B. is supported by the UK Engineering and Physical Sciences Research Council (EPSRC) and by Cambridge Quantum Computing Limited (CQCL). E.G. is supported by ESPRC [EP/P510270/1]. L.W. is supported by the Royal Society. We gratefully acknowledge the support of NVIDIA Corporation with the donation of the Titan Xp GPU used for this research. S.S. is supported by the Royal Society, EPSRC, the National Natural Science Foundation of China, and the grant ARO-MURI W911NF17-1-0304 (US DOD, UK MOD and UK EPSRC under the Multidisciplinary University Research Initiative).\\

\appendix

\end{document}